# Experimental Realization of Special-Unitary Operations in Classical Mechanics by Nonadiabatic Evolutions

Congwei Lu[1], Xulong Wang[1], Guancong Ma[1,2*]

[1]Department of Physics, Hong Kong Baptist University, Kowloon Tong, Hong Kong, China

[2]Shenzhen Institute for Research and Continuing Education, Hong Kong Baptist University, Shenzhen 518000, China

[*]Email: phgcma@hkbu.edu.hk

***Abstract.-*** Artificial classical wave systems such as wave crystals and metamaterials have demonstrated promising capabilities in simulating a wide range of quantum mechanical phenomena. Yet some gaps between quantum and classical worlds are generally considered fundamental and difficult to bridge. Dynamics obeying special unitary groups, e.g., electronic spins described by $\mathrm{SU}(2)$, color symmetries of fundamental particles described by $\mathrm{SU}(3)$, are such examples. In this Letter, we present the experimental realization of universal $\mathrm{SU}(2)$ and $\mathrm{SU}(3)$ dynamic operations in classical mechanical oscillator systems with temporally modulated coupling terms. Our approach relies on the sequential execution of non-adiabatic holonomic evolutions, which are typically used in constructing quantum-logic gates. The method is swift and purely geometric and can be extended to realize more sophisticated dynamic operations. Our results open a new way for studying and simulating quantum phenomena in classical systems.

***Introduction.-*** Physical phenomena of the microscopic world are rich, diverse, and complicated. These phenomena are sometimes very challenging to observe and manipulate experimentally. Therefore, quantum simulations, which emulate complex quantum phenomena using more easily controllable quantum systems, play an important role in modern quantum physics [1,2]. For example, the lattice gauge theories can be mapped onto a Hamiltonian of the qubit system [3–7]. Recent studies have found that such simulations can be performed using operations on qutrits, which have a higher error tolerance compared to qubits [8]. These simulations all require precise $\mathrm{SU(N)}$ control of quantum units (qubits, qutrits, etc.) On a different frontier, classical artificial systems have been successful in emulating phenomena thought to be exclusively quantum mechanical [9,10]. Notable examples include quantum Hall edge states in photonic crystals [11], the topological mechanical modes in isostatic lattices that are related by topological band theory [12], and non-Abelian braiding of light and sound [13,14]. However, quantum and classical physics are described by distinct mathematical laws. One well-known difference is that classical equations of motion are purely real, whereas the process of canonical quantization introduces the imaginary unit into the momentum operator, making the quantum dynamical equations complex [15]. Another profound difference emerges through Lie groups, which play an important role in the theoretical formulation of dynamics of both classical and quantum worlds. A well-known example is the application of the $\mathrm{SO}(3)$ group in describing the three-dimensional (3D) classical rotation of rigid objects [16]: elements of $\mathrm{SO}(3)$ are real matrices that rotate $3 \times 1$ vectors. Thus, the generators of $\mathrm{SO}(3)$ are the angular momentum operators that follow the famous commutator relation $[L_x, L_y] = L_z$. In contrast, many quantum dynamics are characterized by $\mathrm{SU(N)}$ groups. For example, spin-1/2 particles are underpinned by $\mathrm{SU}(2)$, whose generators are the Pauli matrices. The application of $\mathrm{SU}(3)$ has led to the eightfold way for classifying hadrons in quantum chromodynamics [17].

An important question then emerges: can the seemingly fundamental gap between classical $\mathrm{SO}$ and quantum $\mathrm{SU}$ dynamics be bridged? In this work, we successfully realized universal $\mathrm{SU}(2)$ and $\mathrm{SU}(3)$ operations in classical systems. (Here, we consider evolutions of single-particle quantum systems [18–20]. Owing to the absence of particle characteristics in classical systems, there is no direct way to replicate many-body Fock spaces in classical systems.) We apply the method of nonadiabatic evolution, a route to universal quantum logics [21–23], for controlling the dynamics of coupled spring-mass oscillators. This method encodes the $\mathrm{SU(N)}$ evolution in a subspace of the larger system. By controlling the nonadiabatic holonomic evolution of the states in the subspace, the $\mathrm{SU(N)}$ operations are realized by temporally modulating the coupling between multiple oscillators. The protocol is a purely geometric evolution of the system that does not accumulate any dynamic phase. Meanwhile, the nonadiabatic characteristic permits swift operations, which significantly reduce the impact of dissipation and noise.

***Realization of complex coupling.-*** First, we emulate complex entries in the $\mathrm{SU}$ groups. Considering the equations of $N$ coupled classical oscillators, the equations of motion are

$$\ddot{X}_i = -\omega^2 X_i - \sqrt{2\omega} \sum_j \left[ H_{ij} \left( \sqrt{\frac{\omega}{2}} X_j + \mathrm{i} \sqrt{\frac{1}{2\omega}} \dot{X}_j \right) + \mathrm{c.c.} \right], \tag{1}$$

where $X_i = -\omega^2 x_i$ with $x_i$ is the displacement from the equilibrium position of the $i$-th oscillator. $\omega$ is the natural angular frequency of all the oscillators. $H_{ij}$ is the coupling between the $i$-th and $j$-th oscillators. This is a set of purely real second-order differential equations with respect to time. The eigenmodes of mechanical oscillators with a coefficient matrix $H$ are the same as the eigenstates of the time-independent Schrödinger equation with Hamiltonian $H$ [24]. However, if we want to recast it to the form of time-dependent Schrödinger equation, we need to introduce the classical complex variables $\alpha_i \equiv \sqrt{\frac{\omega}{2}} X_i + \mathrm{i} \sqrt{\frac{1}{2\omega}} \dot{X}_i$ [25]. Equation (1) then becomes $\mathrm{i}\dot{\alpha}_i = \omega \alpha_i + \sum_j (H_{ij}\alpha_j + H_{ij}^* \alpha_j^*)$. Then, we enforce the condition $\omega \gg |H_{ij}|$. Therefore, the complex variables $\alpha_i$ can be considered as oscillating near $\omega$. We introduce a slowly varying quantity $\tilde{\alpha}_i = e^{-\mathrm{i}\omega t} \alpha_i$, which gives $\mathrm{i}\dot{\tilde{\alpha}}_i = \sum_j (H_{ij} \tilde{\alpha}_j + H_{ij}^* e^{-2\mathrm{i}\omega t} \tilde{\alpha}_j^*)$, and then apply the rotating wave approximation to neglect the last term with rapid oscillation. Back to $\alpha_i$, we obtain $\mathrm{i}\dot{\alpha}_i \approx \omega \alpha_i + \sum_j H_{ij} \alpha_j$, which has the form of the Schrödinger equation, and $H$ become the Hamiltonian. Note that $H$ is complex only when the oscillators are coupled through velocities. Such interactions are not common in classical systems, perhaps with the exception of the Lorentz force and noninertial forces. Therefore, in our experiments, the coupling is facilitated by active control, as will be described later.

***Synthesis of $SU(2)$ operations.-*** Next, we find an appropriate Hamiltonian to implement SU operations. We begin by considering an $(N+1)$-dim Hamiltonian,

$$H(t) = \Omega(t)\begin{pmatrix} 0_{N\times N} & |\boldsymbol{v}\rangle \\ \langle \boldsymbol{v}| & 0 \end{pmatrix} = \Omega(t)M, \tag{2}$$

where $|\boldsymbol{v}\rangle$ is a normalized vector, i.e., $|\boldsymbol{v}\rangle = (v_1, v_2, \ldots, v_N)^{\mathrm{T}}$, and $\langle \boldsymbol{v}| = |\boldsymbol{v}\rangle^{\dagger}$. And $\Omega(t)$ is a global time modulation. This Hamiltonian has sublattice symmetry: subsystem A has $N$ oscillators that couple to subsystem B with one oscillator [Fig. 1(b)]. Our proposal is to initialize the input at Hilbert space of subsystem A, denoted $\mathcal{H}_A$, then design an appropriate evolution path so that the final state returns to $\mathcal{H}_A$ and achieves the target operation in $\mathcal{H}_A$. This process is schematically shown in Fig. 1(a). Note that the initial and final states are generically not eigenstates of the Hamiltonian. The transition between different eigenspaces of $H(t)$ is what makes the evolution nonadiabatic. Take a three-site system as an example. $\mathcal{H}_A$ is two-dimensional. We can represent the projection of the state in $\mathcal{H}_A$ using the Bloch sphere. At the initial and final moments, the state is located on the surface of the Bloch sphere, while at intermediate moments, the state is inside the Bloch sphere. This means that the state evolves in the full Hilbert space $\mathcal{H}_A \oplus \mathcal{H}_B$ and its projection in $\mathcal{H}_A$ is less than 1. The method of achieving target states by introducing auxiliary systems is widely used in the field of quantum simulation. A typical example is the construction of the direct product space of the system and auxiliary qubits, which allows for the non-Hermitian evolution of the system through post-selection [26,27].

Now we find an evolution path such that the state vectors return to $\mathcal{H}_A$ after a period T. Such an evolution is described by $U_{\mathrm{T}} \coloneqq U(t=\mathrm{T}) = \mathcal{T}\exp\left[-\mathrm{i}\int_0^{t=\mathrm{T}} dt'\, H(t')\right]$. Then we have $U_{\mathrm{T}} = \boldsymbol{I} + M^2\cos\alpha - M^2 - \mathrm{i}M\sin\alpha$, where $\alpha = \int_0^{\mathrm{T}} dt\,\Omega(t)$, and $M^2 = \begin{pmatrix} |\boldsymbol{v}\rangle\langle\boldsymbol{v}| & 0 \\ 0 & 1 \end{pmatrix}$. To evolve the final state back to $\mathcal{H}_A$, we set $\alpha = \pi$ such that only $M^2$ terms appear in $U_{\mathrm{T}}$, i.e.,

$$U_{\mathrm{T}} = \begin{pmatrix} \boldsymbol{I}_{N\times N} - 2|\boldsymbol{v}\rangle\langle\boldsymbol{v}| & 0_{N\times 1} \\ 0_{1\times N} & -1 \end{pmatrix}. \tag{3}$$

This is a unit evolution. The evolution operator in $\mathcal{H}_A$ is $U_A = \boldsymbol{I}_{N\times N} - 2|\boldsymbol{v}\rangle\langle\boldsymbol{v}|$. Let $|\psi(0)\rangle$ and $|\psi'(0)\rangle$ be two arbitrary initial states belonging to $\mathcal{H}_A$. The intermediate states evolve from $|\psi(0)\rangle$ and $|\psi'(0)\rangle$ are $|\psi(t)\rangle$ and $|\psi'(t)\rangle$ respectively. Then, the expectation values of the Hamiltonian over any possible intermediate states and the energy transition between any intermediate states are zero during the whole evolution, i.e., $\langle\psi'(t)|H(t)|\psi(t)\rangle = \langle\psi'(0)|U(t)^{\dagger}H(t)U(t)|\psi(0)\rangle = \langle\psi'(0)|H(t)|\psi(0)\rangle = 0$. This ensures no dynamical contribution appears in the evolution of subsystem A [28–30]. Since there is no requirement for the intermediate states to remain an eigenstate of $H(t)$, $U_{\mathrm{T}}$ is non-adiabatic and holonomic in character. Consequently, the evolution can be significantly speeded up compared to adiabatic ones. Our goal is to achieve the target operation by designing $L$ steps of purely geometric unit evolutions $U_{\mathrm{T}}$.

Now we perform an $\mathrm{SU}(2)$ rotation in $\mathcal{H}_A$, we invoke Eq. (3) becomes $U_{\mathrm{SU}(2)} = \begin{pmatrix} e^{\mathrm{i}\boldsymbol{\sigma}\cdot\boldsymbol{e}\varphi} & 0 \\ 0 & 1 \end{pmatrix}$, where $\boldsymbol{\sigma}$ is the Pauli matrix and $e^{\mathrm{i}\boldsymbol{\sigma}\cdot\boldsymbol{e}\varphi}$ represents a clockwise rotation by an angle $2\varphi$ around $\boldsymbol{e}$-axis on the Bloch sphere. For simplicity, we let $\Omega(t) = \pi/\mathrm{T}$. Using the mathematical equality $e^{\mathrm{i}\boldsymbol{\sigma}\cdot\boldsymbol{e}\varphi} = (\boldsymbol{n}\cdot\boldsymbol{\sigma})(\boldsymbol{n}'\cdot\boldsymbol{\sigma}) = \boldsymbol{n}\cdot\boldsymbol{n}' + \mathrm{i}\boldsymbol{\sigma}\cdot(\boldsymbol{n}\times\boldsymbol{n}') = |\boldsymbol{n}||\boldsymbol{n}'|\cos\varphi + |\boldsymbol{n}||\boldsymbol{n}'|\sin\varphi\,\boldsymbol{\sigma}\cdot\boldsymbol{e}$ with $\boldsymbol{n}(\theta,\phi) = (\sin\theta\cos\phi, \sin\theta\sin\phi, \cos\theta)$, $\boldsymbol{e} = \frac{\boldsymbol{n}\times\boldsymbol{n}'}{|\boldsymbol{n}\times\boldsymbol{n}'|}$ and $\varphi = \arccos(\boldsymbol{n}\cdot\boldsymbol{n}')$ [28], we can obtain

$$U_{\mathrm{SU(2)}} = U_{\mathrm{T}_2}(\boldsymbol{n})U_{\mathrm{T}_1}(\boldsymbol{n}'), \text{ with } U_{\mathrm{T}}(\boldsymbol{n}) = \begin{pmatrix} \boldsymbol{n}\cdot\boldsymbol{\sigma} & 0 \\ 0 & -1 \end{pmatrix}. \tag{4}$$

So the coupling vector in Eq. (2) takes the form $|\boldsymbol{v}\rangle = \left(\sin\frac{\theta}{2}e^{-\mathrm{i}\phi}, -\cos\frac{\theta}{2}\right)^{\mathrm{T}}$. This $\mathrm{SU}(2)$ rotation can be realized by two consecutive unit evolutions. For a generic operation $R \in \mathrm{SU}(2)$, which comprises by three sequential rotations generated by $\sigma_y$ and $\sigma_z$: $R(\alpha,\beta,\gamma) = e^{\mathrm{i}\sigma_z(\beta-\gamma)/2}e^{\mathrm{i}\sigma_y\alpha/2}e^{\mathrm{i}\sigma_z(\beta+\gamma)/2}$ [31], can be realized with at most six such unit evolutions.

Here, we demonstrate the operation $e^{i\frac{\pi}{2}\sigma_z}$. The initial state is chosen as $|\psi(0)\rangle = (1,0,0)^{\mathrm{T}}$. From Eqs. (3,4), we have $e^{i\frac{\pi}{2}\sigma_z} = U_{\mathrm{T}_2}\left[\boldsymbol{n}\left(\frac{\pi}{2},0\right)\right]U_{\mathrm{T}_1}\left[\boldsymbol{n}\left(\frac{\pi}{2},\frac{\pi}{2}\right)\right]$, and $|\boldsymbol{v}\rangle_{\mathrm{T}_1} = -\frac{1}{\sqrt{2}}(\mathrm{i},1)^{\mathrm{T}}$, $|\boldsymbol{v}\rangle_{\mathrm{T}_2} = \frac{1}{\sqrt{2}}(1,-1)^{\mathrm{T}}$. Recalling that $\Omega(t) = \frac{\pi}{\mathrm{T}}$, the entries of the Hamiltonian as functions of time are obtainable, as shown in Fig. 1(c). The wavefunction evolution is shown in Fig. 1(d). Clearly, the end state is $|\psi(2T)\rangle = (\mathrm{i},0,0)^{\mathrm{T}}$. The projection of wave function onto subsystem A, i.e., $|\Psi(\mathrm{t})\rangle = \mathcal{P}_A|\psi(t)\rangle$, can be mapped onto the Bloch sphere, with coordinates given by $X = \langle\Psi(t)|\sigma_x|\Psi(t)\rangle$, as shown in Fig. 1(e). Note that, except for the three instants $0, \mathrm{T}$, and $2\mathrm{T}$, the intermediate states are indeed inside the Bloch sphere.

***Experimental realization of $SU(2)$ operation.-*** We design three coupled oscillators and precisely control the coupling between them by actively feedback-controlled system (see End Matter), such that oscillators 1 and 2 achieve two $\mathrm{SU}(2)$ rotations, $e^{i\frac{\pi}{2}\sigma_z}$ and $e^{i\frac{\pi}{4}\sigma_y}$. The initial state, $|\psi(0)\rangle = (1,0,0)^{\mathrm{T}}$, is prepared by setting an initial angle for oscillator-1 at time $t = 0$. For $e^{i\frac{\pi}{2}\sigma_z}$, the evolutions of all three oscillators under our non-adiabatic control are shown in Fig. 2(a-c). After undergoing two steps of unit evolutions, the state transitions are $(1,0,0)^{\mathrm{T}} \to (0,\mathrm{i},0)^{\mathrm{T}} \to (\mathrm{i},0,0)^{\mathrm{T}}$. The states at time $t = 0, \mathrm{T}, \text{and } 2\mathrm{T}$ are shown in the right panels of Fig. 2(a-c), respectively. The target state $(\mathrm{i},0,0)^{\mathrm{T}}$ featuring maximum (zero) instantaneous angular velocity (displacement) at oscillator-1, and a near-stationary status for oscillators-2 and 3. After time $2\mathrm{T}$, oscillator-1 evolves at its natural frequency. For $e^{i\frac{\pi}{4}\sigma_y}$, the evolutions of all three oscillators are shown in Fig. 2(d-f). The state transitions $(1,0,0)^{\mathrm{T}} \to (1,1,0)^{\mathrm{T}}/\sqrt{2} \to (1,-1,0)^{\mathrm{T}}/\sqrt{2}$ are shown in Fig. 2(d-f), respectively. The target state $(1,-1,0)^{\mathrm{T}}/\sqrt{2}$ featuring opposite (zero) instantaneous angular displacement (velocity) at oscillators-1 and 2, and a near-stationary status for oscillator-3. The period of one unit evolution is $\mathrm{T} = 0.954$ s.

***Realizing $SU(3)$ operations by Householder decomposition.-*** Our protocol can be extended to realize $\mathrm{SU}(3)$ operations. Observe that any element $g \in \mathrm{SU}(3)$ can be described by eight real parameters: $g(\alpha_1,\beta_1,\gamma_1,\alpha_2,\beta_2,\gamma_2,\alpha_3,\beta_3,\gamma_3) = R_{23}(\alpha_1,\beta_1,\gamma_1)R_{12}(\alpha_2,\beta_2,\gamma_2)R_{23}(\alpha_3,\beta_3,\gamma_3)$ [32], where $R_{ij}(\alpha,\beta,\gamma)$ is the $\mathrm{SU}(2)$ operator on the Hilbert space $\mathcal{H}_i \oplus \mathcal{H}_j$ of site-$i$ and site-$j$. Each $\mathrm{SU}(2)$ operation can be achieved by 6 unit evolutions, so an $\mathrm{SU}(3)$ operation requires a total of 18 unit evolutions to complete. Such a long sequence of non-adiabatic controls is challenging in experiments, owing to the presence of dissipation and experimental errors. We mitigate this issue by observing that the evolution operation for subsystem A, $U_A = \boldsymbol{I}_{3\times3} - 2|\boldsymbol{v}\rangle\langle\boldsymbol{v}|$ is a Householder matrix [33]. An $N$-dimensional unitary matrix $\mathrm{U}(N)$ can be expressed as a product of $N-1$ Householder matrices $U_A(\boldsymbol{v}_n), n = 1,2,\cdots,N-1,$ and a phase gate $\Phi(\phi_1,\phi_2,\cdots,\phi_N) = \mathrm{diag}(e^{i\phi_1}, e^{i\phi_2},\cdots,e^{i\phi_N})$ [34]

$$U = U_A(\boldsymbol{v}_1)U_A(\boldsymbol{v}_2)\cdots U_A(\boldsymbol{v}_{N-1})\Phi(\phi_1,\phi_2,\cdots,\phi_N). \tag{5}$$

The phase gate $\Phi(\phi_1,\phi_2,\cdots,\phi_N)$ can be achieved by $N-1$ rotations generated by $\sigma_z$, i.e., sequentially performing the $\mathrm{SU}(2)$ rotation $e^{\mathrm{i}\sigma_z\phi}$ on the Hilbert space $\mathcal{H}_i \oplus \mathcal{H}_{i+1}$ $(1 \le i \le N-1)$. As such, completing a universal $\mathrm{U}(N)$ operation requires at most $3(N-1)$ unit evolutions using the Householder decomposition [Eq. (5)]. So, a generic $\mathrm{SU}(3)$ operator is decomposable into six unit operations. (An experimental implementation of an $\mathrm{SU}(2)$ operation based on Householder matrix decomposition is provided in Supplemental Material [30].)

We experimentally implement a simple case of $\mathrm{SU}(3)$ operation,

$$G = \begin{pmatrix} 0 & -\mathrm{i} & 0 \\ 0 & 0 & 1 \\ \mathrm{i} & 0 & 0 \end{pmatrix}, \tag{6}$$

which can be decomposed as

$$G = U_A(\boldsymbol{v}_1)U_A(\boldsymbol{v}_2)\Phi(0,0,0), \tag{7}$$

with $|\boldsymbol{v}_1\rangle = \frac{1}{\sqrt{2}}(-1,0,\mathrm{i})^{\mathrm{T}}$, $|\boldsymbol{v}_2\rangle = \frac{1}{\sqrt{2}}(0,-1,1)^{\mathrm{T}}$. Because $\Phi(0,0,0)$ is an identity, $G$ only requires two unit evolutions. We can design a system with four coupled oscillators and achieve $G$ for oscillators 1,2, and 3 by precisely modulated the coupling terms. In Fig. 3(c), we show the evolution of the angular displacements $\theta$ and velocities $\dot{\theta}$ of four oscillators from initial state $|2\rangle = (0,1,0,0)^{\mathrm{T}}$. The state transitions are $(0,1,0,0)^{\mathrm{T}} \to (0,0,1,0)^{\mathrm{T}} \to (-\mathrm{i},0,0,0)^{\mathrm{T}}$, which are shown in right side of Fig. 3(c). The results of the time evolution for the other initial states $|1\rangle = (1,0,0,0)^{\mathrm{T}}$ and $|3\rangle = (0,0,1,0)^{\mathrm{T}}$ are provided in the Supplemental Materials . The experimentally obtained $G$ matrix is shown in Figs. 3(a, b).

***Conclusion.-*** We demonstrated non-adiabatic protocols for realizing $\mathrm{SU}(2)$ and $\mathrm{SU}(3)$ dynamics in classical mechanics. It can be further extended to an $(N+1)$-site system to implement the $\mathrm{SU}(N)$ operations. In ref. [30], we present the realization of the Fourier transform operation for qutrits. Future works may aim at extending the system in real space to form a lattice and simulate the non-Abelian lattice gauge fields [35]. Our experimental platform also holds promises for studying quantum nonlinear physics with chiral symmetry [41] by constructing a force that is second-order correlated with velocity and displacement. We anticipate our results will vastly advance the classical simulation of quantum systems.

***Acknowledgments.-***This work was supported by the National Key R&D Program (2022YFA1404403), the Hong Kong Research Grants Council (RFS2223-2S01, 12301822), the National Natural Science Foundation of China (12472088), and the Hong Kong Baptist University (RC-RSRG/23-24/SCI/01, RC-SFCRG/23-24/R2/SCI/12).

*Data availability—The data that support the findings of this Letter are openly available [42].*

## End Matter

***The experimental setup.–*** The system consists of harmonic oscillators with one rotational degree of freedom. Each oscillator consists of an arm with loaded weights, two brushless DC motors (motor A and motor B), tensioned springs connecting the arm to the anchors, and angular displacement measurement electronics, as shown in Fig. 4(a). All oscillators are tuned to the same natural angular frequency $\omega = 2\pi \times 5.24\,\mathrm{rad \cdot s^{-1}}$. Two magnets (indicated by yellow arrows) are fixed to motor A and motor B respectively via shafts, as shown in Fig. 4 (d). Motor A and motor B are fixed to the arm, and they rotate together. The two magnetic rotary position sensors (AMS AS5047P) (indicated by a red arrow) are fixed on the table to measure the rotation of the magnets in real-time, obtaining the angular displacement and velocity of the arm. These signals are sent to a microcontroller and fed back to the two motors, enabling active control of the coupling. The system is capable of programmable and active control of the system parameters, which have been successfully used in several previous experiments [37–41].

Figure 4 (b) shows a schematic diagram of coupled oscillators implementing $\mathrm{SU}(2)$ and $\mathrm{SU}(3)$. The $\mathrm{SU}(2)$ operation utilizes only oscillators-1, 2, and 3 out of the four oscillators. The oscillators-1 and 2 form the operational space $\mathcal{H}_A$. The $\mathrm{SU}(3)$ operation requires all four oscillators. Oscillators-1, 2, and 3 form the operational space $\mathcal{H}_A$. Here, we use site-2 and site-4 as examples to demonstrate how to apply time-varying coupling. As shown in Fig. 4 (c), microcontroller B (Espressif ESP32) is programmed to read the instantaneous angular displacement $\theta_4$ and velocity $\dot{\theta}_4$ of the arm of site-4 (indicated by a green arrow). Then Microcontroller B generates a signal based on $\theta_4$ and velocity $\dot{\theta}_4$ to the driver circuit (STMicroelectronics L6234PD) to instruct the motor B to exert an active torque $\tau_i(t) = a_{ij}\theta_j(t) + b_{ij}\dot{\theta}_j(t)$, where the complex hopping term $\mathrm{Re}(H_{24}) \propto a_{24}$ and $\mathrm{Im}(H_{24}) \propto b_{24}$. Since the Hamiltonian is Hermitian, the arm of site-4 needs to be subjected to a torque that depends on the angular displacement and velocity of the arm of site-2. The proportional coefficients are $a_{42} = a_{24}$ and $b_{42} = -b_{24}$, respectively. To ensure the proper observation of the evolution, it is desirable to minimize the system loss. This is also achieved by active control: microcontroller A controls motor A to apply a torque to the arm that is proportional to the angular velocity, i.e., $\eta_i(t) = c_i\dot{\theta}_i(t)$. This compensates for the dissipation of each oscillator.

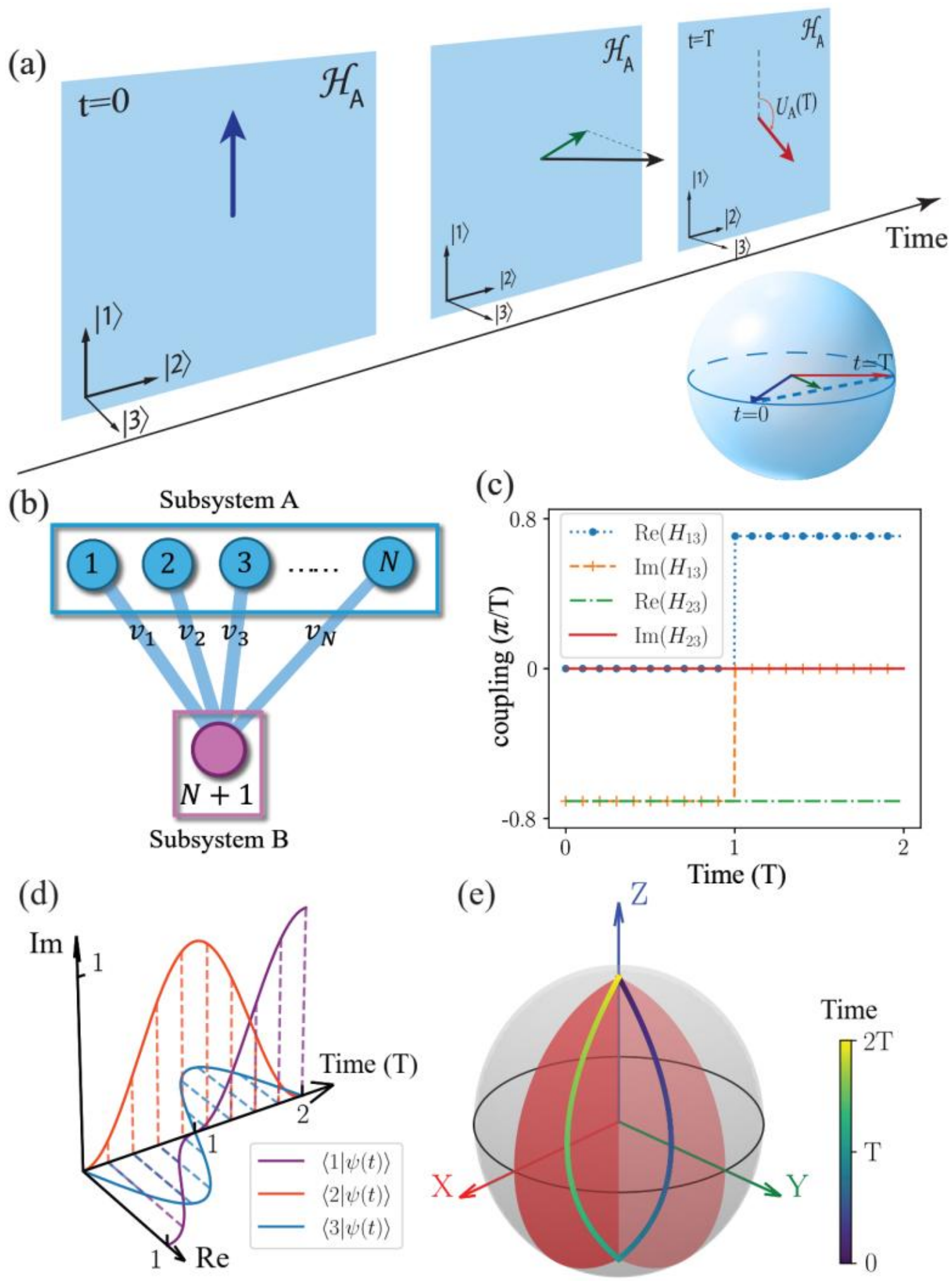

FIG. 1. (a) Non-adiabatic control drives an initial state in Hilbert subspace $\mathcal{H}_A$ to evolve to an end state in $\mathcal{H}_A$ by a unitary operation $U_A(\mathrm{T})$. The evolution can be mapped onto the Bloch sphere defined over $\mathcal{H}_A$. (b) An $(N+1)$-site system consisting of two subsystems A (*N* sites) and B (1 site). (c-e) The evolution of the coupling vector $|\boldsymbol{v}\rangle = \mathrm{T}/\pi(H_{13}, H_{23})^{\mathrm{T}}$ (c) and the wavefunction (d) during the implementation of $e^{i\frac{\pi}{2}\sigma_z}$. (e) The projection of the wavefunction to the Bloch sphere of Hilbert subspace $\mathcal{H}_A$.

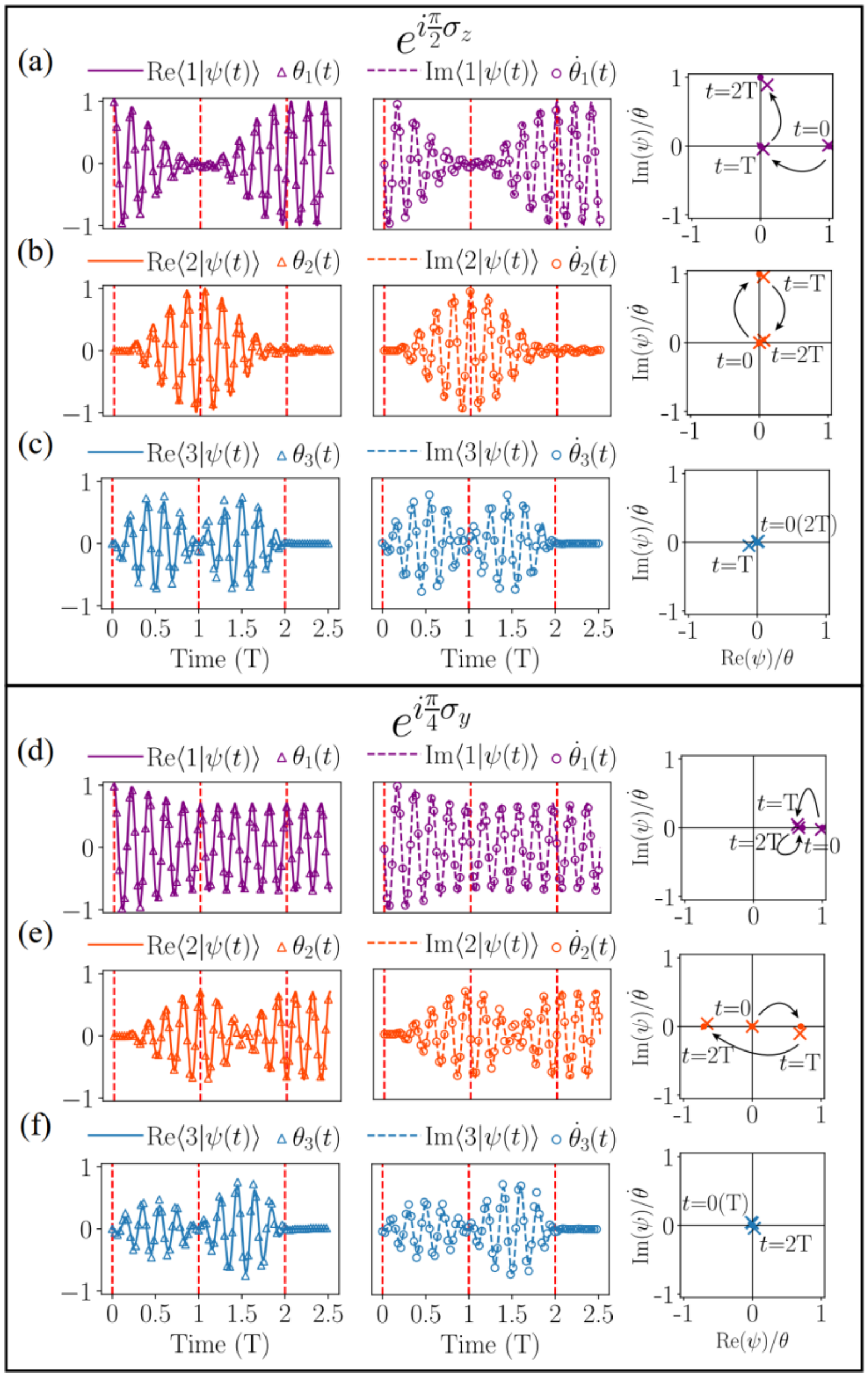

FIG. 2. The experimental results of SU(2) dynamics. (a-c) plot the operation $e^{i\frac{\pi}{2}\sigma_z}$, (d-f) plot $e^{i\frac{\pi}{4}\sigma_y}$. The real (imaginary) part of the wavefunctions, which corresponds to the rotational angular $\theta$ (angular velocity $\dot{\theta}$). The curves are theoretical results computed using experimental parameters. The red vertical dashed lines mark the moments $t = 0, \mathrm{T}$, and $2\mathrm{T}$. The states at these moments are shown on the right side, where the crosses and points represent the experimental and theoretical results, respectively.

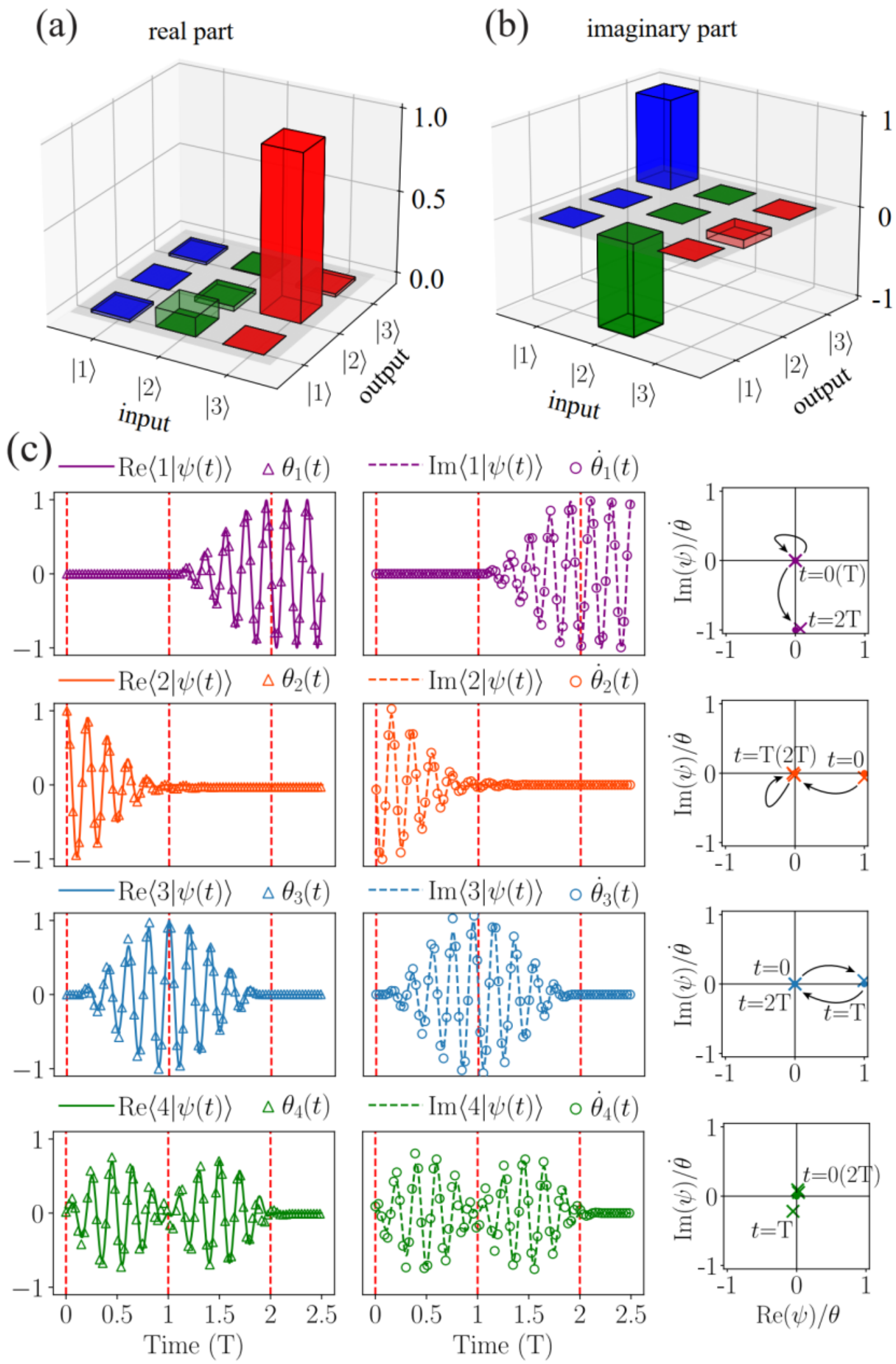

FIG. 3. (a-b) The experimental outputs of the operator $G \in \mathrm{SU}(3)$ [Eq. (6)]. (c) The measured time-dependent wavefunction. Right panels: the states at times $t = 0, \mathrm{T}$, and $2\mathrm{T}$ from experimental and theoretical results.

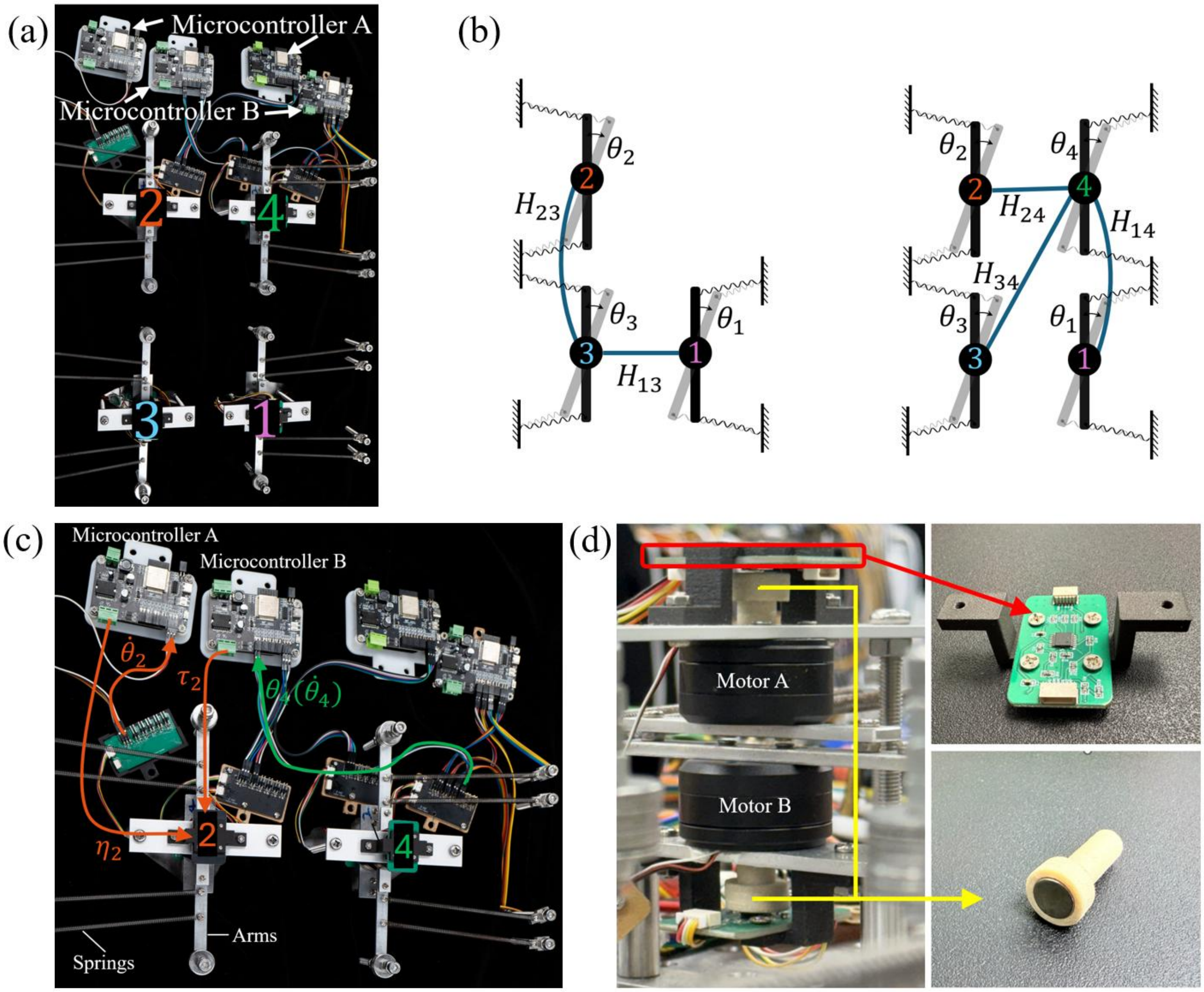

FIG. 4. (a) A photograph of the coupled oscillators. (b) The schematic of the four-oscillator system. Three (four) oscillators are used for SU(2) (SU(3)) operations. (c) Photos of the coupled rotational oscillators. (d) The angular displacement measurement setup. The front end of the shaft is embedded with a magnet (indicated by yellow arrows), and the other end is fixed to a motor, rotating together with the motor and the arm. The change in the magnetic field caused by the magnet's rotation is detected by the magnetic rotary position sensor (indicated by a red arrow).

# Supplemental Materials

# Experimental Realization of Special-Unitary Operations in Classical Mechanics by Nonadiabatic Evolutions

Congwei Lu[1], Xulong Wang[1], Guancong Ma[1,2*]

[1]Department of Physics, Hong Kong Baptist University, Kowloon Tong, Hong Kong, China
[2]Shenzhen Institute for Research and Continuing Education, Hong Kong Baptist University, Shenzhen 518000, China
[*]Email: phgcma@hkbu.edu.hk

**Contents**

## I. The geometric nature of the non-adiabatic holonomic evolution

In this section, we explain why a non-adiabatic unit evolution in the main text is purely geometric, i.e., there is no dynamical-phase effect. We consider a system described by an $M$-dimensional Hilbert space $\mathcal{H}_M$ with Hamiltonian $H(t)$. We define an $L$-dimensional subspace $\mathcal{H}_L(t)$ (computation space) spanned by a set of instantaneous orthogonal basis vectors $\{|\psi_k(t)\rangle\}_{k=1}^{L}$, each vector $|\psi_k(t)\rangle$ satisfying the Schrödinger equation

$$\mathrm{i}\partial_t|\psi_k(t)\rangle = H(t)|\psi_k(t)\rangle. \tag{S1}$$

The unit evolution $U_\mathrm{T} \coloneqq U(t=\mathrm{T}) = \mathcal{T}\exp\left[-\mathrm{i}\int_0^{t=\mathrm{T}} dt'\, H(t')\right]$ is a holonomy matrix acting on $\mathcal{H}_L$ if $|\psi_k(t)\rangle$ satisfies the following conditions [1]

$$\text{(i)}\quad \sum_{k=1}^{L}|\psi_k(\mathrm{T})\rangle\langle\psi_k(\mathrm{T})| = \sum_{k=1}^{L}|\psi_k(0)\rangle\langle\psi_k(0)|,$$

$$\text{(ii)}\quad \langle\psi_k(t)|H(t)|\psi_l(t)\rangle = 0,\ \ k,l = 1,2,\cdots,L. \tag{S2}$$

*Proof.*

We introduce a set of bases $\{|\nu_k(t)\rangle\}_{k=1}^{L}$ of $\mathcal{H}_L(t)$. $|\psi_k(t)\rangle$ can be expressed as

$$|\psi_k(t)\rangle = \sum_{l=1}^{L}|\nu_l(t)\rangle C_{lk}(t), \tag{S3}$$

and we can properly choose $C_{lk}(t)$ such that condition (ii) be satisfied,

$$|\nu_k(\mathrm{T})\rangle = |\nu_k(0)\rangle = |\psi_k(0)\rangle,\ \ k,l = 1,2,\cdots,L. \tag{S4}$$

Then

$$|\psi_k(\mathrm{T})\rangle = \sum_{l=1}^{L}|\nu_l(\mathrm{T})\rangle C_{lk}(\mathrm{T}) = \sum_{l=1}^{L}|\psi_l(0)\rangle C_{lk}(\mathrm{T}). \tag{S5}$$

It is obvious that $C(\mathrm{T}) = U_\mathrm{T}$. Substituting Eq. (S3) into the Schrödinger equation Eq. (S1) yields

$$\frac{d}{dt}C_{lk}(t) = \mathrm{i}\sum_{m=1}^{L}[A_{lm}(t) - K_{lm}(t)]C_{mk}(t), \tag{S6}$$

where $A_{kl}(t) = \mathrm{i}\langle\nu_k(t)|\frac{d}{dt}|\nu_l(t)\rangle$, and $K_{kl}(t) = \langle\nu_k(t)|H(t)|\nu_l(t)\rangle$. If condition (ii) is satisfied, then

$$\langle\psi_m(t)|H(t)|\psi_m(t)\rangle = \sum_{l,k=1}^{L} K_{kl}(t)C_{km}^*(t)C_{lm}(t) = 0. \tag{S7}$$

Therefore, $K_{kl}(t) = 0$, so

$$U_{\mathrm{T}} = C(\mathrm{T}) = \mathcal{T}\exp\left[-\mathrm{i}\int_0^{\mathrm{T}} dt\, A(t)\right]. \tag{S8}$$

In the main text, the Hilbert space $\mathcal{H}_A$ of subsystem A corresponds to the subspace $\mathcal{H}_L$. The unit evolution in Eq. (3) in the main text is a linear transformation of the subspace $\mathcal{H}_L$, thus it obviously satisfies condition (i). On the other hand, $\langle\psi_k(t)|H(t)|\psi_l(t)\rangle = \langle\psi_k(0)|U(t)^\dagger H(t)U(t)|\psi_l(0)\rangle = \langle\psi_k(0)|H(t)|\psi_l(0)\rangle = 0$, so condition (ii) is also satisfied. Therefore, any unit evolution is a holonomic evolution, with the dynamical contribution being zero.

## II. Householder matrix decomposition

In this section, we show that any $U(N)$ matrix can be expressed as a product of $(N-1)$ Householder matrices $U_A(\boldsymbol{v}_n) = \boldsymbol{I}_{N\times N} - 2|\boldsymbol{v}_n\rangle\langle\boldsymbol{v}_n|, n = 1,2,\cdots,N-1$, and a phase matrix $\Phi(\phi_1,\phi_2,\cdots,\phi_N) = \mathrm{diag}(e^{i\phi_1},e^{i\phi_2},\cdots,e^{i\phi_N})$ [2],

$$U = U_A(\boldsymbol{v}_1)U_A(\boldsymbol{v}_2)\cdots U_A(\boldsymbol{v}_{N-1})\Phi(\phi_1,\phi_2,\cdots,\phi_N). \tag{S9}$$

*Proof.*

First, we define

$$|\boldsymbol{v}_1\rangle = \frac{|\boldsymbol{u}_1\rangle - e^{\mathrm{i}\phi_1}|\boldsymbol{e}_1\rangle}{\sqrt{2(1-|U_{11}|)}}, \tag{S10}$$

where $|\boldsymbol{u}_n\rangle = [U_{1n},U_{2n},\cdots,U_{Nn}]^{\mathrm{T}}$, $\phi_1 = \arg U_{11}$ and $|\boldsymbol{e}_1\rangle = [1,0,\cdots,0]^{\mathrm{T}}$. Then we can verify that

$$\begin{aligned} &U_A(\boldsymbol{v}_1)|\boldsymbol{u}_1\rangle = e^{\mathrm{i}\phi_1}|\boldsymbol{e}_1\rangle,\\ &U_A(\boldsymbol{v}_1)|\boldsymbol{u}_n\rangle = |\boldsymbol{u}_n\rangle + e^{-\mathrm{i}\phi_2}U_{1n}\frac{|\boldsymbol{u}_1\rangle - e^{\mathrm{i}\phi_1}|\boldsymbol{e}_1\rangle}{1-|U_{11}|}, n \geq 2 \end{aligned} \tag{S11}$$

and

$$\langle\boldsymbol{e}_1|U_A(\boldsymbol{v}_1)|\boldsymbol{u}_n\rangle = 0 \quad (n = 2,3,\cdots,N). \tag{S12}$$

Therefore

$$U_A(\boldsymbol{v}_1)U = \begin{pmatrix} e^{\mathrm{i}\phi_1} & 0 & \cdots & 0 \\ 0 & & & \\ \vdots & & U_{N-1} & \\ 0 & & & \end{pmatrix}, \tag{S13}$$

where $U_{N-1}$ is a $U(N-1)$ matrix. Then we repeat above procedure on $U_A(\boldsymbol{v}_1)U$. We define

$$|\boldsymbol{v}_2\rangle = \frac{|\boldsymbol{u}_2'\rangle - e^{\mathrm{i}\phi_2}|\boldsymbol{e}_2\rangle}{\sqrt{2(1-|(U_A(\boldsymbol{v}_1)U)_{22}|)}}, \tag{S14}$$

where $|\boldsymbol{u}_2'\rangle = [(U_A(\boldsymbol{v}_1)U)_{1n}, (U_A(\boldsymbol{v}_1)U)_{2n}, \cdots, (U_A(\boldsymbol{v}_1)U)_{Nn}]^{\mathrm{T}}$, i.e., second column of $U_A(\boldsymbol{v}_1)U, \phi_2 = \arg[U_A(\boldsymbol{v}_1)U]_{22}$ and $|\boldsymbol{e}_2\rangle = [0,1,0,\cdots,0]^{\mathrm{T}}$. Then we can verify that

$$\begin{aligned} &U_A(\boldsymbol{v}_2)|\boldsymbol{u}_1'\rangle = |\boldsymbol{u}_1'\rangle, \\ &U_A(\boldsymbol{v}_2)|\boldsymbol{u}_2'\rangle = e^{\mathrm{i}\phi_2}|\boldsymbol{e}_2\rangle, \\ &U_A(\boldsymbol{v}_2)|\boldsymbol{u}_n'\rangle = |\boldsymbol{u}_n'\rangle + e^{-\mathrm{i}\phi_2}(U_A(\boldsymbol{v}_1)U)_{2n}\frac{|\boldsymbol{u}_2'\rangle - e^{\mathrm{i}\phi_2}|\boldsymbol{e}_2\rangle}{1-|(U_A(\boldsymbol{v}_1)U)_{22}|}, n \geq 3 \end{aligned} \tag{S15}$$

and

$$\begin{aligned} \langle\boldsymbol{e}_1|U_A(\boldsymbol{v}_2)|\boldsymbol{u}_n'\rangle &= e^{\mathrm{i}\phi_1}\delta_{1,n} \\ \langle\boldsymbol{e}_2|U_A(\boldsymbol{v}_2)|\boldsymbol{u}_n'\rangle &= e^{\mathrm{i}\phi_2}\delta_{2,n} \end{aligned} \tag{S16}$$

Therefore

$$U_A(\boldsymbol{v}_2)U_A(\boldsymbol{v}_1)U = \begin{pmatrix} e^{\mathrm{i}\phi_1} & 0 & 0 & \cdots & 0 \\ 0 & e^{\mathrm{i}\phi_2} & 0 & \cdots & 0 \\ 0 & 0 & & & \\ \vdots & \vdots & & U_{N-1} & \\ 0 & 0 & & & \end{pmatrix}. \tag{S17}$$

By repeating the same procedure $N-1$ times, we can obtain

$$U_A(\boldsymbol{v}_{N-1})\cdots U_A(\boldsymbol{v}_1)U = \Phi(\phi_1,\phi_2,\cdots,\phi_N). \tag{S18}$$

Since $U_A(\boldsymbol{v})$ is Hermitian and Unitary, i.e., $U_A(\boldsymbol{v}) = U_A(\boldsymbol{v})^{-1} = U_A(\boldsymbol{v})^{\dagger}$, we have

$$U = U_A(\boldsymbol{v}_1)U_A(\boldsymbol{v}_2)\cdots U_A(\boldsymbol{v}_{N-1})\Phi(\phi_1,\phi_2,\cdots,\phi_N). \tag{S19}$$

□

The phase gate can be achieved by applying rotations generated by $\sigma_z$ on each of the $(N-1)$ pairs of sites. The final result may differ from $\Phi(\phi_1,\phi_2,\cdots,\phi_N)$ only by an overall phase factor.

## III. The isomorphic mapping between $\mathrm{SU}(2)$ and $\mathrm{SO}(3)$.

In this section, we explain why does $e^{\mathrm{i}\boldsymbol{\sigma}\cdot\boldsymbol{e}\varphi}$ represent a rotation angle of $2\varphi$. The doubled rotation angle is fundamentally due to the fact that $\mathrm{SU}(2)$ is a double cover of $\mathrm{SO}(3)$. This can be seen by mapping a two-component complex state vector $|\psi\rangle$ to a unit vector $\boldsymbol{r}$ in a Euclidean space,

$$|\psi\rangle \to \boldsymbol{r} = (X,Y,Z)^T = \left(\langle\psi|\sigma_x|\psi\rangle, \langle\psi|\sigma_y|\psi\rangle, \langle\psi|\sigma_z|\psi\rangle\right)^T.$$

Next, we consider a change of coordinates $\boldsymbol{r} \to \boldsymbol{r}'$, when $|\psi\rangle$ changed by the operation $e^{\mathrm{i}\boldsymbol{\sigma}\cdot\boldsymbol{e}\varphi}$, i.e., $|\psi\rangle \to |\psi'\rangle = e^{\mathrm{i}\boldsymbol{\sigma}\cdot\boldsymbol{e}\varphi}|\psi\rangle$. Without loss of generality, we prove the case of rotation around the z-axis,

$$|\psi\rangle \to |\psi'\rangle = e^{\mathrm{i}\sigma_z\varphi}|\psi\rangle.$$

Then the image is changed by

$$\boldsymbol{r} \to \boldsymbol{r}' = \begin{pmatrix} \langle\psi|e^{-\mathrm{i}\sigma_z\varphi}\sigma_x e^{\mathrm{i}\sigma_z\varphi}|\psi\rangle \\ \langle\psi|e^{-\mathrm{i}\sigma_z\varphi}\sigma_y e^{\mathrm{i}\sigma_z\varphi}|\psi\rangle \\ \langle\psi|\sigma_z|\psi\rangle \end{pmatrix} = \begin{pmatrix} \langle\psi|\cos(2\varphi)\,\sigma_x + \sin(2\varphi)\,\sigma_y|\psi\rangle \\ \langle\psi|\cos(2\varphi)\,\sigma_y - \sin(2\varphi)\,\sigma_x|\psi\rangle \\ \langle\psi|\sigma_z|\psi\rangle \end{pmatrix}$$

$$= \begin{pmatrix} \cos(2\varphi) & \sin(2\varphi) & 0 \\ -\sin(2\varphi) & \cos(2\varphi) & 0 \\ 0 & 0 & 1 \end{pmatrix} \begin{pmatrix} X \\ Y \\ Z \end{pmatrix}$$

This represents a unit vector in Euclidean space rotating clockwise around the z-axis by $2\varphi$. Rotations around the other two coordinate axes and any arbitrary axis can be derived through the same calculations. When $\varphi = \pi$, the corresponding operation $e^{\mathrm{i}\boldsymbol{\sigma}\cdot\boldsymbol{e}\pi}$ changes $|\psi\rangle$ to $-|\psi\rangle$, while the corresponding $\boldsymbol{r}$ rotates $2\pi$ around the $\boldsymbol{e}$-axis, leaving its state unchanged.

## IV. Realization of an $\mathrm{SU}(2)$ operator based on Householder matrix decomposition

In this section, we give a proof-of-principle experiment to demonstrate how to use Householder matrix decomposition to implement $\mathrm{SU}(2)$ operations. We experimentally implement a simple case of $\mathrm{SU}(2)$ operation,

$$e^{-\mathrm{i}\frac{\pi}{4}\sigma_y} = \frac{\sqrt{2}}{2}\begin{pmatrix} 1 & -1 \\ 1 & 1 \end{pmatrix}, \tag{S20}$$

which can be decomposed by Householder matrix,

$$e^{-\mathrm{i}\frac{\pi}{4}\sigma_y} = U_A(\boldsymbol{v}_1)\Phi(0,\pi) = -\mathrm{i}U_A(\boldsymbol{v}_1)\Phi\left(\frac{\pi}{2},-\frac{\pi}{2}\right). \tag{S21}$$

From the first equality of Eq. (S21), we find that the result obtained using the Householder matrix decomposition is consistent with the result obtained using the decomposition $e^{-\mathrm{i}\frac{\pi}{4}\sigma_y} = (\boldsymbol{n}\cdot\boldsymbol{\sigma})(\boldsymbol{n}'\cdot\boldsymbol{\sigma})$ in the main text, where $\boldsymbol{n} = \boldsymbol{n}(\theta,\phi) = (\sin\theta\cos\phi, \sin\theta\sin\phi, \cos\theta)$. It can be directly verified that $\boldsymbol{n}\cdot\boldsymbol{\sigma} = U_A(\boldsymbol{v}_1)$ and $\boldsymbol{n}'\cdot\boldsymbol{\sigma} = \Phi(0,\pi)$, where $\boldsymbol{n} = \boldsymbol{n}\left(\frac{\pi}{4},0\right)$ and $\boldsymbol{n}' = \boldsymbol{n}(0,0)$ respectively. However, the decomposition of the second equality is more general. For an arbitrary phase gate $\Phi(\phi_1,\phi_2)$, it may not be represented as $\boldsymbol{n}\cdot\boldsymbol{\sigma}$, but it can be represented as $e^{\frac{-\mathrm{i}(\phi_1+\phi_2)}{2}}\Phi\left(\frac{\phi_2-\phi_1}{2},\frac{\phi_1-\phi_2}{2}\right)$, i.e., a rotation operation around the z-axis multiplied by an overall phase. As a proof of principle, we experimentally realize $e^{-\mathrm{i}\frac{\pi}{4}\sigma_y}$ by implementing the operation $K = U_A(v_1)\Phi\left(\frac{\pi}{2},-\frac{\pi}{2}\right)$ in the second equality of Eq. (S21). The phase gate $\Phi\left(\frac{\pi}{2},-\frac{\pi}{2}\right) = e^{i\frac{\pi}{2}\sigma_z} =$ $U_{\mathrm{T}_2}\left[\boldsymbol{n}\left(\frac{\pi}{2},0\right)\right]U_{\mathrm{T}_1}\left[\boldsymbol{n}\left(\frac{\pi}{2},\frac{\pi}{2}\right)\right]$ requires two unit evolutions, which are determined by $|\boldsymbol{v}\rangle_{\mathrm{T}_1} = -\frac{1}{\sqrt{2}}(\mathrm{i},1)^{\mathrm{T}}$, $|\boldsymbol{v}\rangle_{\mathrm{T}_2} = \frac{1}{\sqrt{2}}(1,-1)^{\mathrm{T}}$. The third unit evolution $U_A(\boldsymbol{v}_1)$ is determined by $|\boldsymbol{v}_1\rangle = \left(\sin\frac{\pi}{8},-\cos\frac{\pi}{8}\right)^{\mathrm{T}}$. The initial state $|\psi(0)\rangle = (-\mathrm{i},0,0)^{\mathrm{T}}$ is prepared by setting a zero displacement and maximum reverse instantaneous velocity for oscillator-1 at time $t=0$. The evolution curves of all three oscillators under our non-adiabatic control are shown in Fig. S1(a-c). Step I and step II are used to complete the phase gate, and step III completes $U_A(\boldsymbol{v}_1)$. The state transitions are $(-\mathrm{i},0,0)^{\mathrm{T}} \rightarrow (0,1,0)^{\mathrm{T}} \rightarrow (1,0,0)^{\mathrm{T}} \rightarrow (1,1,0)^{\mathrm{T}}/\sqrt{2}$. The states at time $t = 0, \mathrm{T}, 2\mathrm{T}$ and $3\mathrm{T}$ are shown in the right panels of Fig. S1(a-c), respectively. The system reaches the end state featuring zero instantaneous angular velocity and the same displacement at oscillators-1 and 2, and a near-stationary status for oscillator-3. After time $3\mathrm{T}$, oscillators-1 and 2 evolve freely at their natural frequencies.

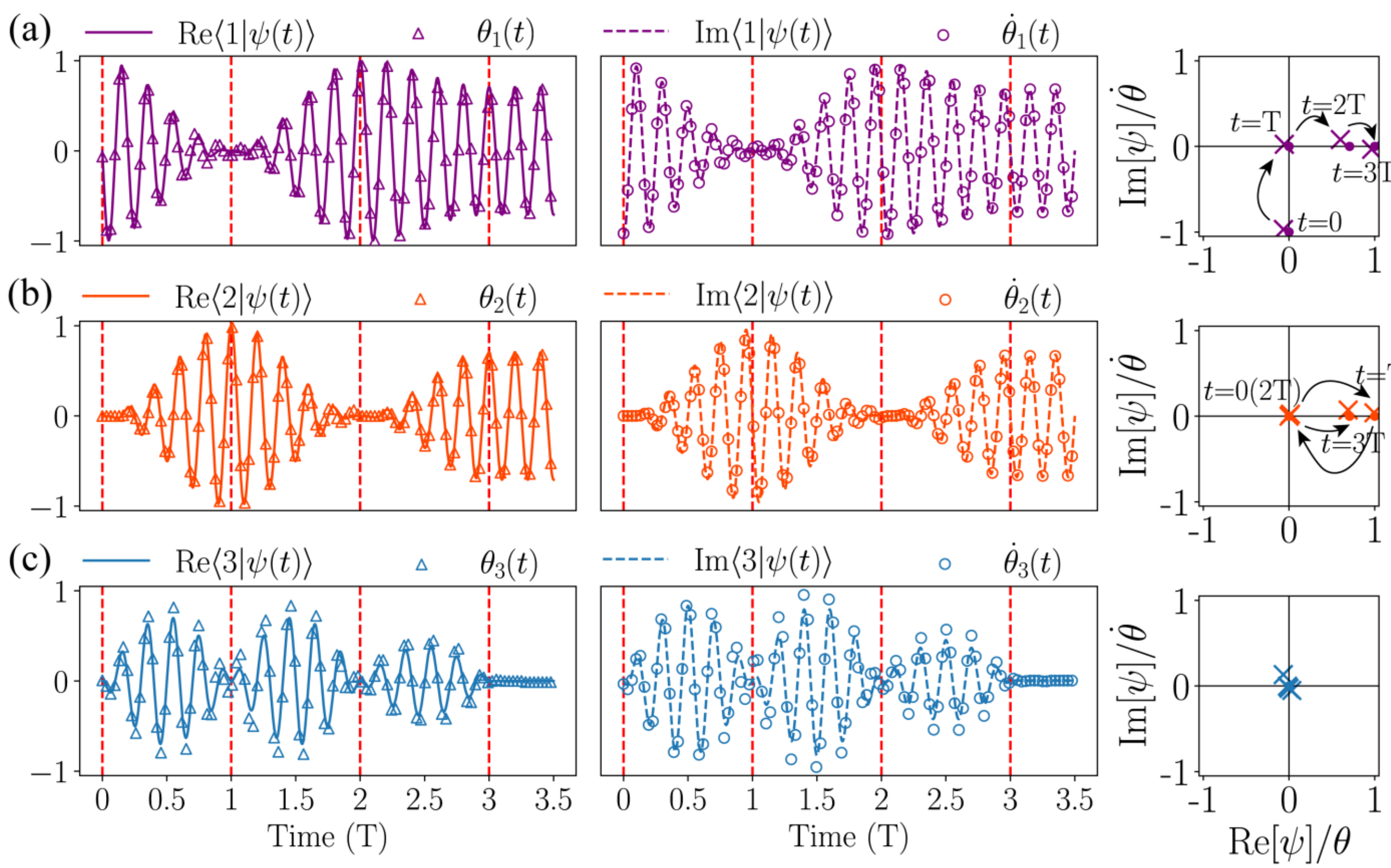

**FIG. S1.** The measured evolution of the wavefunction in a three-oscillator system implementing $e^{-\mathrm{i}\frac{\pi}{4}\sigma_y}$. The red vertical dashed lines mark the moments $t = 0, \mathrm{T}$, $2\mathrm{T}$ and $3\mathrm{T}$. The states at these moments are shown on the right side, where the crosses and points represent the experimental and theoretical results, respectively.

## V. Householder decomposition of an $\mathbf{SU(3)}$ operator

In this section, we present an example of utilizing the Householder matrix decomposition to implement the $\mathrm{SU}(3)$ operation, $G$ matrix introduced in the main text.

As discussed in the main text, there are two methods for decomposing the $\mathrm{SU}(3)$ operation. The first method is to decompose it into three $\mathrm{SU}(2)$ operations, requiring at most 18 unit evolutions. As for the G matrix, it can be decomposed into

$$G = \begin{pmatrix} 0 & -\mathrm{i} & 0 \\ 0 & 0 & 1 \\ \mathrm{i} & 0 & 0 \end{pmatrix} = \begin{pmatrix} 1 & \\ & e^{\frac{\mathrm{i}\pi}{2}\sigma_y} \end{pmatrix} \begin{pmatrix} e^{\frac{\mathrm{i}\pi}{2}\sigma_x} & \\ & 1 \end{pmatrix} \begin{pmatrix} e^{\mathrm{i}\pi\sigma_x} & \\ & 1 \end{pmatrix}. \tag{S22}$$

Therefore, 6 unit evolutions are required. However, under the Householder matrix decomposition,

$$G = U_A(\boldsymbol{v}_1)U_A(\boldsymbol{v}_2)\Phi(0,0,0), \tag{S23}$$

where $|\boldsymbol{v}_1\rangle = \frac{1}{\sqrt{2}}(-1,0,\mathrm{i})^{\mathrm{T}}$, $|\boldsymbol{v}_2\rangle = \frac{1}{\sqrt{2}}(0,-1,1)^{\mathrm{T}}$. And only two unit evolutions are required.

## VI. Full results of the $\mathbf{SU(3)}$ operation $\boldsymbol{G}$

In Fig. S2(a-b), we show the evolutions of four oscillators' angular displacements $\theta$ and velocities $\dot{\theta}$ from different initial states $(1,0,0,0)^{\mathrm{T}}$ and $(0,0,1,0)^{\mathrm{T}}$ respectively. At the key moments $t = 0, \mathrm{T}$, and $2\mathrm{T}$, the state transitions are $(1,0,0,0)^{\mathrm{T}} \to (1,0,0,0)^{\mathrm{T}} \to (0,0,\mathrm{i},0)^{\mathrm{T}}$ and $(0,0,1,0)^{\mathrm{T}} \to (0,1,0,0)^{\mathrm{T}} \to (0,1,0,0)^{\mathrm{T}}$, which are shown in right panels of Fig. S2(a-b).

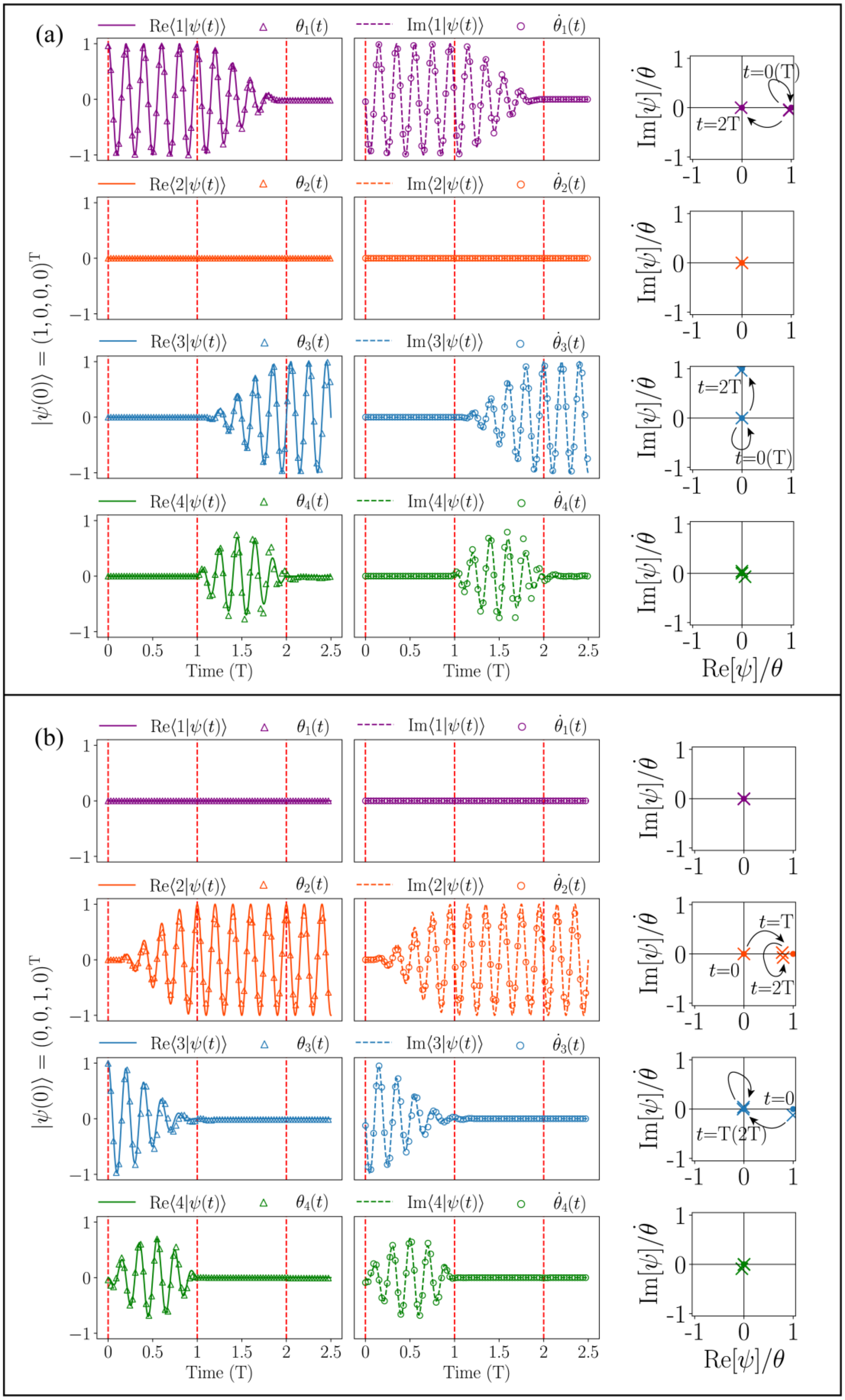
(a)
(b)
$|\psi(0)\rangle = (1,0,0,0)^T$
$|\psi(0)\rangle = (0,0,1,0)^T$
Re⟨1|ψ(t)⟩ θ₁(t) Im⟨1|ψ(t)⟩ θ̇₁(t)
Re⟨2|ψ(t)⟩ θ₂(t) Im⟨2|ψ(t)⟩ θ̇₂(t)
Re⟨3|ψ(t)⟩ θ₃(t) Im⟨3|ψ(t)⟩ θ̇₃(t)
Re⟨4|ψ(t)⟩ θ₄(t) Im⟨4|ψ(t)⟩ θ̇₄(t)
Time (T)
Im[ψ]/θ̇
Re[ψ]/θ
t=0(T)
t=2T
t=0
t=T
t=T(2T)

**FIG. S2.** The evolutions of the wavefunction of the four-oscillator system implementing $G$ are shown for different initial states. The curves are theoretical results, and the markers are obtained from experimental measurements. The red vertical dashed lines mark the moments $t = 0, T$, and $2T$. The states at these moments are shown on the right panels.

## VII. Discrete Fourier transformation of qutrit

In this section, we propose a feasible scheme to implement the discrete Fourier transformation (DFT) of a qutrit using Householder matrix decomposition. The DFT of a qutrit has the following matrix form,

$$F = \frac{1}{\sqrt{3}}\begin{pmatrix} 1 & 1 & 1 \\ 1 & e^{\frac{i2\pi}{3}} & e^{-\frac{i2\pi}{3}} \\ 1 & e^{-\frac{i2\pi}{3}} & e^{\frac{i2\pi}{3}} \end{pmatrix}. \tag{S24}$$

$F$ is a $\mathrm{U}(3)$ matrix but not $\mathrm{SU}(3)$. But it becomes $\mathrm{SU}(3)$ upon the multiplication of a global phase factor $e^{\frac{i\pi}{6}}$. The phased DFT can be decomposed by

$$F' = e^{\frac{i\pi}{6}}F = e^{\frac{i\pi}{6}}U_A(\boldsymbol{v}_1)U_A(\boldsymbol{v}_2)\Phi\left(\frac{\pi}{6}, \frac{5\pi}{12}, -\frac{7\pi}{12}\right), \tag{S25}$$

where $|\boldsymbol{v}_1\rangle = \frac{1}{2}\sqrt{1+\frac{1}{\sqrt{3}}}\left(1-\sqrt{3},1,1\right)^{\mathrm{T}}$, $|\boldsymbol{v}_2\rangle = \sqrt{\frac{1+\sqrt{2}}{2\sqrt{2}}}\left(0,1-\sqrt{2},-\mathrm{i}\right)^{\mathrm{T}}$. And

$$\Phi\left(\frac{\pi}{6}, \frac{5\pi}{12}, -\frac{7\pi}{12}\right) = \begin{pmatrix} 1 & 0 & 0 \\ 0 & e^{\frac{i5\pi}{12}} & 0 \\ 0 & 0 & e^{-\frac{i5\pi}{12}} \end{pmatrix}\begin{pmatrix} e^{\frac{i\pi}{6}} & 0 & 0 \\ 0 & 1 & 0 \\ 0 & 0 & e^{-\frac{i\pi}{6}} \end{pmatrix}. \tag{S26}$$

The DFT requires 6 unit evolutions. The matrix elements of the Hamiltonian change over time as shown in Fig. S3 (a). The evolutions of wave functions under different initial states are shown in Fig. S3(b). After the operation $Fe^{-\frac{i\pi}{6}}$, the target states are $\frac{1}{\sqrt{3}}e^{-\frac{i\pi}{6}}(1,1,1)^{\mathrm{T}}$, $\frac{1}{\sqrt{3}}e^{-\frac{i\pi}{6}}\left(1, e^{\frac{i2\pi}{3}}, e^{-\frac{i2\pi}{3}}\right)^{\mathrm{T}}$ and $\frac{1}{\sqrt{3}}e^{-\frac{i\pi}{6}}\left(1, e^{-\frac{i2\pi}{3}}, e^{\frac{i2\pi}{3}}\right)^{\mathrm{T}}$ respectively. The outputs are shown in Fig. S3(c).

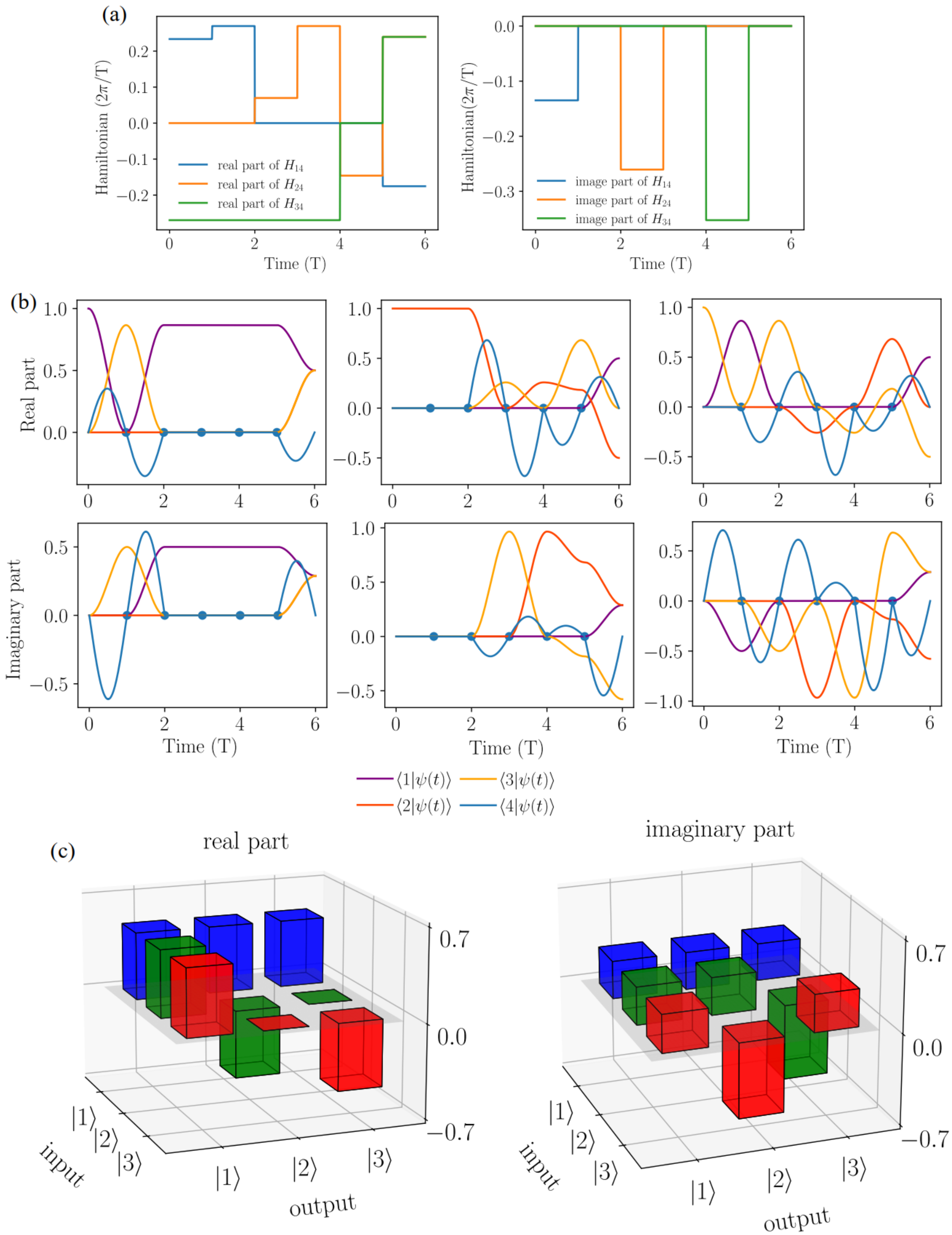

**FIG. S3.** (a) Entries of the Hamiltonian as functions of time. (b) The evolution of the wavefunction implementing operation $F$ for different initial states $|1\rangle = (1,0,0,0)^{\mathrm{T}}$, $|2\rangle = (0,1,0,0)^{\mathrm{T}}$ and $|3\rangle =$

$(0,0,1,0)^{\mathrm{T}}$, respectively. (c) The theoretical output results for different initial states.